\documentclass[fleqn,usenatbib]{mnras}

\usepackage{newtxtext,newtxmath}

\usepackage[T1]{fontenc}

\DeclareRobustCommand{\VAN}[3]{#2}
\let\VANthebibliography\thebibliography
\def\thebibliography{\DeclareRobustCommand{\VAN}[3]{##3}\VANthebibliography}

\usepackage{graphicx}	
\usepackage{amsmath}	

\usepackage{amssymb}	

\defcitealias{Rosotti19}{R19}
\defcitealias{Trapman20}{Tr20}
\defcitealias{T22}{Tr22}
\defcitealias{Toci21}{To21}

\title[Analytic solution for the gas size]{An analytic solution to measure the gas size in protoplanetary discs in the viscous self-similar scenario}

\author[C. Toci et al.]{
Claudia Toci,$^{1,2}$\thanks{E-mail:claudia.toci@unimi.it}
Giuseppe Lodato,$^{1}$
Francesco Gerardo Livio,$^{1}$
Giovanni Rosotti,$^{3,1,4}$
Leon Trapman$^{5}$
\\
$^{1}$Dipartimento di Fisica, Università degli Studi di Milano, Via Giovanni Celoria, 16, 20133, Milano, Italy\\
$^{2}$European Southern Observatory, Karl-Schwarzschild-Strasse 2, D-85748 Garching bei München, Germany\\
$^{3}$School of Physics and Astronomy, University of Leicester, Leicester LE1 7RH, UK\\
$^{4}$Leiden University, Niels Bohrweg 2, NL-2333 CA Leiden, Netherlands \\
$^{5}$Department of Astronomy, University of Wisconsin-Madison, 475 N Charter St, Madison, WI, USA
}

\date{Accepted 2022 October 26. Received 2022 October 25; in original form 2022 August 9}

\pubyear{2022}

\begin{document}
\label{firstpage}
\pagerange{\pageref{firstpage}--\pageref{lastpage}}
\maketitle

\begin{abstract}
In order to understand which mechanism is responsible for accretion in protoplanetary discs, a robust knowledge of the observed disc radius using gas tracers such as  $^{12}$CO and other CO isotopologues is pivotal. Indeed, the two main theories proposed, viscous accretion and wind-driven accretion, predict different time evolution for the disc radii. In this Letter, we present an analytical solution for the evolution of the disc radii in viscously evolving protoplanetary discs using  $^{12}$CO as a tracer, under the assumption that the $^{12}$CO radius is the radius where the surface density of the disc is equal to the threshold for CO photo-dissociation. We discuss the properties of the solution and the limits of its applicability as a simple numerical prescription to evaluate the observed disc radii of populations of discs. Our results suggest that, in addition to photo-dissociation, also freeze out plays an important role in setting the disc size. We find an effective reduction of the CO abundance by about two orders of magnitude at the location of CO photo-dissociation, which however should not be interpreted as the bulk abundance of CO in the disc.
The use of our analytical solution will allow to compute disc sizes for large quantities of models without using expensive computational resources such as radiative transfer calculations. 
\end{abstract}




\section{Introduction}
Protoplanetary discs, reservoirs of material surrounding protostars, are the site of planet formation and early evolution. 
Understanding what are the mechanisms at play in their evolution is therefore fundamental to fully explain the huge variety of exo-planetary systems observed around main-sequence stars \citep{Morbi16}.

Gas and dust stored in discs need to lose angular momentum to fall onto the central star. Since the work of \cite{LBP74}, macroscopic viscosity (parameterised according to the $\alpha$ prescription of \citealt{SS73}) has traditionally been considered responsible for the redistribution of angular momentum within the disc. However, the physical origin of this viscosity (magneto-rotational instabilities, gravitational instabilities, or other mechanisms) is still debated \citep{Balbus03,KL16}. As a consequence of this redistribution of angular momentum, a prediction of the viscous theory is that the physical disc radii should \textit{increase} with time, an effect generally known as viscous spreading.

Recently, new models based on the removal of angular momentum through the interaction of magnetic winds and the disc have been proposed (e.g., \citealt{Bai16,Tabone21,Tabone22_2}). According to this models, the physical disc radius should not evolve with time (\citealt{T22}, hereafter \citetalias{T22}), being either \textit{constant} or \textit{decrease} with time for different disc magnetizations, with the exception of highly magnetized discs, where the disc radius can increase \citep{Bai16}. Therefore, it is in principle possible to discriminate between the two accretion theories through an analysis of the time evolution of the disc radius in different star-forming regions. 

In the literature, the dust or gas radius of discs is usually measured as the radius that contains a certain value (usually the 68$\%$ or 90$\%$) of the disc flux.
However, it is not so straightforward to directly connect the \textit{observed} disc radius using different tracers with the \textit{physical} disc radius, i.e. the radius occupied by the material composing the disc (for a review see \citealt{Miotello22}). First of all, the dust radius measured using ALMA continuum emission (see e.g., \citealt{Ansdell16} for the Lupus star-forming region, \citealt{Long19L} for Taurus, \citealt{Ofiucone} for Ophiucus, \citealt{Barenfeld16B} for Upper Scorpius and \citealt{Hendler20} for an evolutionary study) traces the effect of the radial drift on large dust grains rather then an eventual disc spreading, as large grains, which dominate the opacity of the emission, drift toward the star, reducing the dust disc radii (\citealt{Rosotti19}, hereafter \citetalias{Rosotti19}). Secondly, the gas disc radius (hereafter called $R_{\rm CO}$), commonly measured using the bright rotational line emission of the CO molecule (e.g.,\citealt{Ansdell18} for Lupus), can be interpreted as the radius where CO is photodissociated, rather than the outer edge of the disc (\citealt{Trapman19},\citealt{Trapman20}, hereafter \citetalias{Trapman20}). Practically, this means that, at the radius we are capable to trace, the column density of CO is equal to the threshold value needed for CO molecules to self-shield from external radiation, while outside of this radius all the CO is removed from the disc due to photodissociative effects.

Moreover, as this radius is often located in the outermost part of the disc, it is also not a good tracer for the bulk of the mass and it is very dependent on the external shape of the surface density (\citealt{Toci21}, hereafter \citetalias{Toci21}). As a consequence, the possible presence of the viscous spreading effect is hidden \citep{Manara22}. A combined measurement of the dust and gas radii of populations of protoplanetary discs (for the Lupus star-forming region, see \citealt{Sanchis21}) can partially solve these issues, allowing to distinguish between the two theories, but a better characterisation of the internal structure of the discs and the role of substructures in the disc is needed \citep{Toci21, Apostolos, Zagaria22}. 

In order to perform these demographic studies, the role of the new results coming from surveys of different star forming region is pivotal (for a recent review, see \citealt{Manara22}). Therefore, the need for appropriate theoretical and numerical tools such as population synthesis models, devised to be compared with populations of hundreds of discs, is day by day increasing. Population synthesis models are a powerful way to test evolutionary models (e.g., \citealt{LodatoScardoni,Sellek20, Somigliana20,Zagaria22}) and the effect of different initial conditions \citep{Somigliana22} studying the outcomes of different physical mechanisms on relevant quantities such as dust and gas disc radii.
However, generating and evolving a synthetic populations of discs through numerical methods (and the dust and gas evolution with time) is computationally expensive. It is even more expensive to evaluate the synthetic flux emitted by all the discs at all the timesteps in the dust continuum emission or in the different CO emission lines (to then compute the radius containing the $68\%$ or $90 \%$ of the flux of the models, $R_{\rm{CO,68\%}}$ or $R_{\rm CO, 90\%}$) in an accurate way, i.e., using radiative transfer codes. 
An analytical representation of the sub-mm surface brightness of discs, useful to greatly reduce the numerical cost, is to use a Plank function $B_\nu$, knowing the surface density of the dust component analysed and the opacity of the dust (e.g., \citetalias{Rosotti19}). Currently, a similar analytical representation for the gas radii measured using CO emission lines, $R_{\rm CO}$, is lacking for the viscous scenario, while an approximate treatment for the MHD case has been introduced in \citetalias{T22}.

In this work, we provide an analytical solution for evaluating the value of the disc size $R_{\rm CO}$ of discs in the viscously accreting scenario using $^{12}$CO as a tracer, under the assumption that the $R_{\rm CO}$ radius is defined as the radius where the surface density of the disc is equal to the threshold for $^{12}$CO photo-dissociation \citepalias{Trapman20,T22}. 

The paper is organised as follows. In section~\ref{sec:methods} we recall the 
derivation of the self-similar solutions (sec.~\ref{sec:SelfSim}) and we provide the analytical expression to evaluate $R_{\rm CO}$ (sec~\ref{sec:sol}), in section~\ref{sec:prop} we discuss the properties of the analytical solution and we compare our findings with previous results obtained with numerical codes. We discuss the implication of the comparison in section~\ref{sec:disc} and, finally, in section \ref{sec:conclusions} we draw our conclusions.

\section{Methods}\label{sec:methods}
In the following, we briefly summarise the main results of the viscous theory and we describe our analytical calculation to evaluate gas radii of discs considering  $R_{\rm CO}$ as the radius where the surface density of the disc is equal to the threshold for $^{12}$CO photo-dissociation. 

\subsection{The self-similar solution}\label{sec:SelfSim} 

The surface density of a protoplanetary disc evolved according to the viscous prescription is a function of radius $R$ and time $t$, $\Sigma(R,t)$. A fundamental parameter is the disc viscosity $\nu$, whose origin and exact value, as mentioned above, is still unknown. It is usually described, following \citealt{SS73}, as proportional to the sound speed of the gas in the disc $c_s$ and to the disc thickness, $H = c_s/\Omega$, where $\Omega$ is the Keplerian disc angular velocity, $\nu = \alpha c_s H$.
The value of $\alpha$, the dimensionless scale parameter, previously thought to be $\sim 10^{-2}$ in the literature \citep{Hartmann98}, is more recently expected to be in the range $10^{-3}-10^{-4}$ for most of the discs \citep{Teague16, Flaherty20, Miotello22}.

One analytical solution of the gas viscous evolution equation, often used to describe the disc evolution, is the \textit{self similar solution} \citep{LBP74}. The viscosity is assumed to depend on radius as a simple power-law:
\begin{equation}
    \nu=\nu_c(R/R_c)^{\gamma},
\end{equation}
where $R_c$ is a scale radius, $\nu_c$ is the value of the viscosity at $R_c$ and $\gamma$ is a free index. 
The solution for the surface density $\Sigma$ is then
\begin{equation}\label{eq:self_similar}
    \Sigma(R,t)=\frac{M_0}{2\pi R_c^2}(2-\gamma)\left( \frac{R}{R_c} \right)^{-\gamma}T^{-\eta}
    \exp\left( - \frac{(R/R_c)^{(2-\gamma)}}{T}\right),
\end{equation}
where $M_0$ is the initial ($t=0$) total disc mass, $\eta=(5/2-\gamma)/(2-\gamma)$, $T=1+t/t_\nu$ and the natural time scale of the system, the viscous time $t_\nu$, is defined as
\begin{equation}
    t_\nu=\frac{R_c^2}{3(2-\gamma)^2\nu_c}.
\end{equation}
From eq.~\ref{eq:self_similar}, it can be seen that scale radius $R_c$ is the initial truncation radius of the disc. It evolves with time as
\begin{equation}
 R_c(t)=R_c T^{\frac{1}{2-\gamma}}.   
\end{equation}
The integral of eq.~\ref{eq:self_similar} over radius gives the time evolution of the total disc mass mass,
\begin{equation}
M_d(t) = M_0 T^{1-\eta}.
\end{equation}
As the disc mass has to decrease with time due to the accretion of gas on the star, this implies that $\gamma < 2$.
This condition naturally implies a growing value for $R_{c}(t)>R_c$, meaning that the physical disc size increase with time.
It is possible to analytically calculate also the mass accreted due to the viscous redistribution of angular momentum in the disc, given by:
\begin{equation}
\dot{M} = - \frac{d M}{dt} = (\eta -1)\frac{M_0}{t_\nu}T^{-\eta}.
\end{equation}
For the standard choice $\gamma=1$, we have that $\dot{M} \sim T^{-\eta} \sim t^{-1.5}$. 
\begin{figure*}
    \centering
    \includegraphics[width=\textwidth]{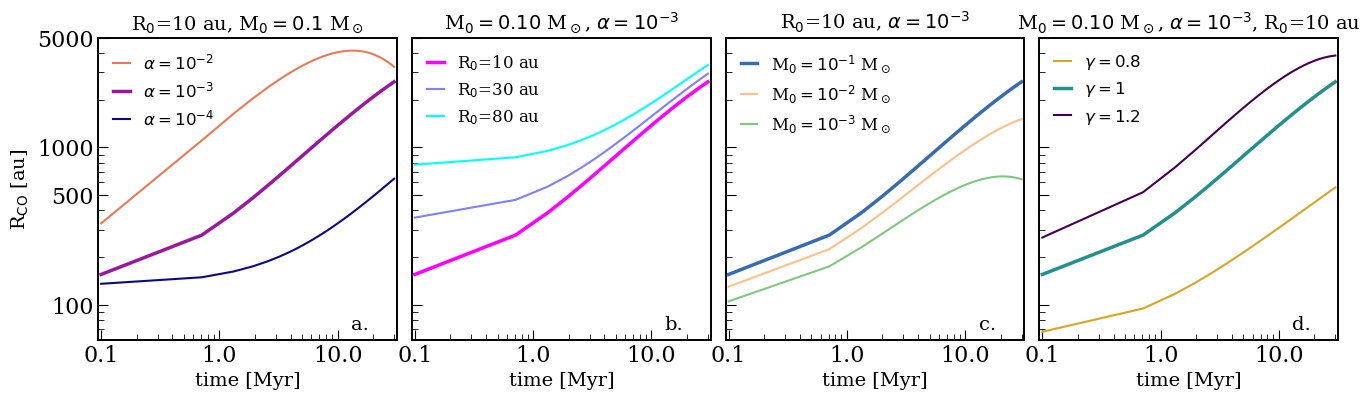}
    \caption{Values of R$_{\rm CO}$ for different initial conditions (both axes are shown in the logarithmic scale). Our reference case is a disc with M$_0=0.1$ M$_\odot$, $R_c=10$ au and $\alpha=10^{-3}$, shown in all the panels with a ticker line. Panel a. shows the time behaviour of the analytical solution for different viscosity values $\alpha=10^{-2},10^{-3},10^{-4}$; Panel b. shows the time behaviour of the analytical solution for different initial scale radii $R_c=10,30,80$ au; Panel c. shows the time behaviour of the analytical solution for different  initial disc masses M$_0=0.001,0.01,0.1$ M$_\odot$; Panel d. shows the time behaviour of the analytical solution for different values of the exponential cut-off $\gamma=0.8,1,1.2$. }
    \label{fig:fig_1}
\end{figure*}

\subsection{An analytical solution for the gas radius $R_{\rm CO}$}\label{sec:sol}

$^{12}$CO rotational emission lines are used to measure the gas radius of a disc either from observations (e.g., \citealt{Barenfeld16B,Ansdell18,Sanchis21}) or models (e.g., \citetalias{Toci21}). 
As discussed by several authors \citepalias{Rosotti19,Toci21,T22}, $R_{\rm CO}$ coincides approximately with the position where the column density of the CO is low enough for it to not self-shield efficiently anymore.
Indeed, the disc evolution leads to a decrease of the disc mass with time, lowering the surface density and allowing CO to be more easily photo-dissociated. This reduces the gaseus CO in the outer part of the disc, and the CO flux from this region is heavily decreased. The value of the $^{12}$CO column density threshold for self-shielding, $\hat{N}_{\rm CO}$, is expected to be of the order of $10^{15}-10^{16}$ cm$^{-2}$ \citep{Ewine88,Facchini17}.  Chemical depletion of CO through grain-surface chemistry only is, for timescales of $\sim$ 10 Myr, negligible when measuring the disc sizes using $^{12}$CO \citepalias{Trapman20}. On the contrary, on time-scales of 1-3 Myr, the combined effect of CO freeze out, grain-surface chemistry and vertical mixing can reduce the CO abundance in the upper layers of the discs up to two orders of magnitude \citep{Kr20}. 
Given a value for the critical column density threshold, the critical surface density of the tracer can be evaluated as
\begin{equation}
    \Sigma_{\rm crit}=\frac{1}{\xi_{\rm CO}}\mu_{\rm gas}\hat{N}_{\rm CO}, 
\end{equation}
where $\xi_{\rm CO}$ is the relative abundance of $^{12}$CO with respect to  H$_2$ in the disc, and $\mu_{\rm gas}$ is the mean molecular weight of the gas in the disc, that we assume equal to the mean molecular weight of H$_2$ molecule\footnote{Note that the standard value for protoplanetary disc is $\mu_{\rm gas} \sim 2.3 \mu_{H}=1.15\mu_{H_2}$. For simplicity and as in \citetalias{Rosotti19} and \citetalias{Toci21} we fix $\mu_{\rm gas}=\mu_{H_2}$.}, $\mu_{\rm gas}=\mu_{H_2}$ . For the standard values of the literature ($\xi_{\rm CO}=10^{-4}, \hat{N}_{\rm CO}=10^{16}$ cm$^{-2}$), $\Sigma_{\rm crit} \sim 1.7\times 10^{-4}$ g cm$^{-2}$. 

It is then possible, under these assumptions, to analytically determine the value of $R_{\rm CO}$ for discs and its time evolution in the viscous scenario. To do so we solved the equation
\begin{equation}\label{eq:rco}
    \Sigma(R,t)=\Sigma_0(2-\gamma)\left( \frac{R}{R_c} \right)^{-\gamma}T^{-\eta}
    \exp\left( - \frac{(R/R_c)^{(2-\gamma)}}{T}\right)=\Sigma_{\rm crit},
\end{equation}
where $\Sigma_0=M_0/(2\pi R_c^2)$ is the initial surface density of the disc. Evaluating the logarithms of both terms, after some algebra the equation reduces to
\begin{equation}
\gamma R_c^{2-\gamma}\ln R+\frac{R^{2-\gamma}}{T}=R_c^{2-\gamma}\ln\left( \frac{\Sigma_0}{\Sigma_{\rm crit}}\frac{(2-\gamma)R_c^{\gamma}}{T^{\eta}}  \right).\label{eq:rc}
\end{equation}
Solving for $R$, the solution of eq.\ref{eq:rc} is given by 
\begin{equation}
    R_{\rm CO}(t)=R_c\left[ \frac{\gamma}{2-\gamma}T(t)\mathcal{W}\left( \frac{(2-\gamma)^{\frac{2}{\gamma}}}{\gamma}\Phi^{\frac{2-\gamma}{\gamma}}T(t)^{-\frac{5}{2\gamma} }    \right)\right], \label{eq:sol}
\end{equation}
where we defined $\Phi=\Sigma_0/\Sigma_{\rm crit}$ and $\mathcal{W}(x)$ is a multivalued mathematical function called \textit{Lambert function} (for an analysis of the properties see e.g., \citealt{Lambert}). For $\gamma=1$ the solution is
\begin{equation}
    R_{\rm CO}(t)=R_c\left[ T(t)\mathcal{W}\left( \Phi T(t)^{-\frac{5}{2}}    \right)\right]. \label{eq:sol_gamma_1}
\end{equation}
It is natural to ask if the value of $R_{\rm CO}$ will grow indefinitely or if it will reach a maximum value and then decrease, and on which timescale. Indeed, for larger $t_\nu$ and low mass discs, the rapid accretion of material from the disc to the star allows all the outer parts of the disc to reach the threshold for photo-dissociation of $^{12}$CO. While the disc expands due to viscous spreading, an increasingly larger part of the disc attains a surface density below the photo-dissociation threshold, leading to an \textit{apparent} decrease of the observed disc size with time.

We analytically determine the time when the expansion of $R_{\rm CO}$ stops and the trend of the function inverts from increasing to decreasing, that we call "inversion time" and we mark as $t_{\rm inv}$, by determining where the maximum of the solution eq.\ref{eq:sol} is:
\begin{equation}\label{eq:inv_time} 
  t_{\rm inv}=t_\nu\left[ \left(\frac{2-\gamma}{\gamma}\right)^{\frac{4}{5}}
  \left(\frac{5-2\gamma}{2 \gamma}\right)^{-\frac{2\gamma}{5}}
  \Phi^{\frac{2(2-\gamma)}{5}}e^{{\frac{2\gamma-5}{5}}} -1
  \right],
\end{equation}
which reduces for $\gamma=1$ to 
\begin{equation}t_{\rm inv}=t_\nu[e^{-3/5}(2/3\Phi)^{2/5}-1]\sim t_\nu(0.47\Phi^{2/5}-1). 
\end{equation}
Naturally, $t_{\rm inv}$ depends from the evolutionary parameters of the disc. For a Solar Mass star\footnote{fixing the stellar mass determines also the temperature profile and H/R}, $\gamma=1$ and $\Phi >> 1$ (equal to $\Sigma>>\Sigma_{\rm crit}$, true for a reasonable range of M$_0$, $R_c$ and $\alpha$ for a Solar Mass star), eq.~\ref{eq:inv_time} can be also written as:
\begin{equation}\label{eq:inv_time_reduced}
t_{\rm inv}\sim 140 \rm{Myr} \left[ \left( \frac{\Sigma_{\rm crit}}{1.7\cdot10^{-4}~{\rm g~cm}^2} \right)^{\frac{-2}{5}}
\left( \frac{\alpha}{10^{-3}}\right)^{-1} \left( \frac{R_c}{10 ~\rm{au}} \right)^{\frac{6}{5}}\left(\frac{M_0}{0.1 M_\odot}\right)^{\frac{2}{5}}  \right].
\end{equation}
In general, a population of discs is expected to survive (and be observable) for 5 - 10 Myr, according to the stellar mass distribution and the physical mechanisms responsible for their evolution (see e.g., \citealt{Manara22}).
Eq.~\ref{eq:inv_time_reduced} shows that $t_{\rm inv} > 10$ Myr for most of the values of the parameters (unless $\alpha$ is exceptionally high, $\alpha\sim 10^{-2}$, or the density of the disc is extremely low, $\Phi \sim$ 1-100, in agreement with \citetalias{Trapman20}). Thus, in the viscous scenario,  the observed disc size of a population should always increase with time. In general, recalling that $\gamma < 2$ \citep{LodatoScardoni}, we can conclude that $t_{\rm inv}$ is larger for lower values of $\alpha$ or larger values of M$_0$ and $R_c$ for all the $\gamma$.

The analytical solution for  $R_{\rm CO}$ shown in eq. \ref{eq:sol} can be applied also to other optically thick molecules with abundances determined by a photo-dissociation threshold. The only dependence on the tracer can be found in the values of $\Sigma_{\rm crit}$ and $\xi_{\rm CO}$. In the case of other CO isotopologues, if the CO abundance in the outer part of the disc is also governed by the self-shielding, the above solution still holds, but the specific value of $\Sigma_{\rm crit}$ and $\xi_{\rm CO}$ vary.

\section{Properties of the solution}\label{sec:prop}

To measure the gas size of discs using $^{12}$CO we assume standard values, fixing $\xi_{\rm CO}=10^{-4}$, $\hat{N}_{\rm CO}=10^{16}$ cm$^{-2}$, a value for $H/R$ at 10 au of 0.033 and the mass of the star to 1 Solar Mass (1 M$_\odot$) and we evaluate the time evolution of  $R_{\rm CO}$ from 0.1 to 30 Myr. Note that we do not expect to observe any disc-bearing young stars after $\sim$ 10 Myr due to the disc removal processes, but we evolved our analytic solution to a long age to show the general features of the analytical solution and check the presence of a possible maximum.  We show our results in Fig.~\ref{fig:fig_1}. 
Our reference case is a disc with $\gamma=1$, M$_0=0.1$ M$_\odot$, $R_c=10$ au and $\alpha=10^{-3}$. The values of the parameters of this model are chosen to be representatives of the intermediate values found in the literature\footnote{For a detailed explanation of the parameters selection see e.g, \citetalias{Rosotti19} or \citetalias{Toci21}}. 
We then vary $\alpha$ between $10^{-2}$ and $10^{-4}$, $R_c$ between 10 and 80 au, M$_0$ between 0.001 and 0.1 M$_\odot$ and $\gamma$ between 0.8 and 1.2, in order to span the parameter space.  

As expected, the trend of the solutions matches what was found in previous works. In particular:
\begin{itemize}
\item Fixing all the other parameters, the observed disc size $R_{\rm CO}$ is larger for more viscous discs  (Fig.~\ref{fig:fig_1}.a), for initially larger discs (Fig.~\ref{fig:fig_1}.b), for more massive discs (Fig.~\ref{fig:fig_1}.c) and for larger values of $\gamma$ (Fig.~\ref{fig:fig_1}.d) \citepalias{Toci21}.  Fig.~\ref{fig:fig_1}.b also show that the values of $R_{\rm CO}$ obtained for discs with different $R_c$, initially different, are almost the same after a few Myr, pointing out that the initial value of $R_c$ plays a minor role in determining the value of $R_{\rm CO}$ on later stages. 
\item The inversion time $t_{\rm inv}$ is reached only for extreme cases before 10 Myr (very high viscosity or very low mass discs). This implies that, neglecting all the other evolutionary processes such as internal and external photo-dissociation, $R_{\rm CO}$ should always increase with time during the average lifetime of a disc if it is evolving due to viscous processes \citepalias{Trapman20}.
\item A different value for $\hat{N}_{\rm CO}$ modifies where $^{12}$CO is removed. From an observational point of view, the measure is degenerate with the uncertainty of the initial CO abundance and the disc temperature, thus it is very difficult to constrain. Using the analytic solution, we probed the effect of $\hat{N}_{\rm CO}=10^{15}$ cm$^{-2}$ \citep{Ewine88}: having a lower $\hat{N}_{\rm CO}$ value leads to a factor $\sim$ 1.2 larger gas radii (in analytical and numerical models). This effect is more significant for larger values of the viscosity, for higher $^{12}$CO abundances and with time. 
\item Just by inspecting the range of values of the different panels of Fig.~\ref{fig:fig_1}, it can be seen that the $R_{\rm CO}$ values are larger than the one commonly measured in the literature for populations of discs (e.g, \citealt{Ansdell18} for Lupus or \citealt{Long22} for Taurus). Indeed, values of $R_{\rm CO}$ obtained analytically are always $>400$ au at 2-3 Myr while results obtained observationally or with numerical codes predict values $\sim$ 100-300 au.
\end{itemize}
To probe the discrepancy between the analytical and numerical predictions, we provide  in the following section a quantitative comparison between the values of $R_{\rm CO}$ analytically predicted and those obtained with numerical codes from \citetalias{Toci21}.
\begin{figure}
    \centering
    \includegraphics[width=0.75\columnwidth]{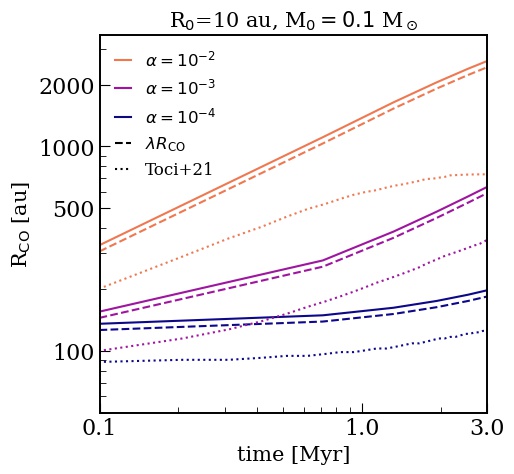}
    \caption{Values of R$_{\rm CO}$ (both axis are shown in logarithmic scale) for discs with M$_0=0.1$ M$_\odot$, $R_c=10$ au and $\alpha=10^{-2},10^{-3},10^{-4}$. Quantities evaluated analytically are shown with solid lines, dashed lines are the analytical values corrected with the coefficient found in \citet{Trapman19}, and dotted lines are the values of R$_{\rm CO,90\%}$ numerically found in \citetalias{Toci21}.}
    \label{fig:fig_2}
\end{figure}
\begin{figure*}
    \centering
    \includegraphics[width=\textwidth]{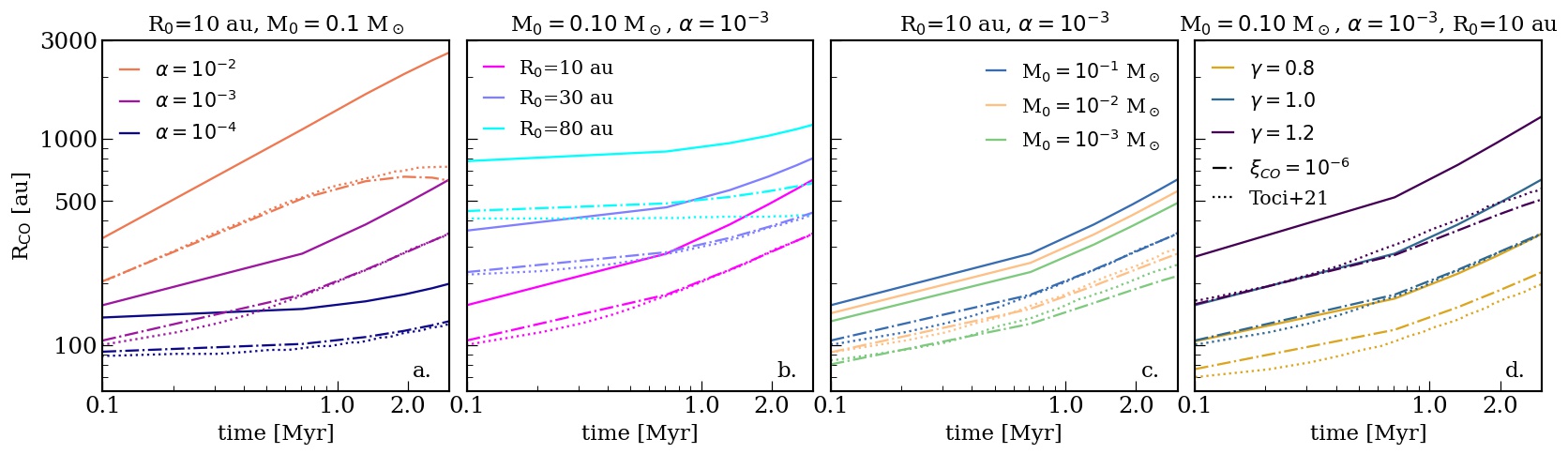}
    \caption{Values of R$_{\rm CO}$ from Fig.\ref{fig:fig_1} with $\xi_{\rm CO}=10^{-4}$ and values of R$_{\rm CO}$ for the same disc parameters but with $\xi_{\rm CO}=10^{-6}$, shown as solid and dash-dotted lines respectively (both axis are shown in logarithmic scale). For comparison, values of R$_{\rm CO,90\%}$ obtained in \citetalias{Toci21} with the same disc parameters are shown as dotted lines. Panel a. shows the time behaviour of the analytical solution for different viscosity values $\alpha=10^{-2},10^{-3},10^{-4}$; Panel b. shows the time behaviour of the analytical solution for different initial scale radii $R_c=10,30,80$ au; Panel c. shows the time behaviour of the analytical solution for different initial disc masses M$_0=0.001,0.01,0.1$ M$_\odot$; Panel d. shows the time behaviour of the analytical solution for different values of the exponential cut-off $\gamma=0.8,1,1.2$. }
    \label{fig:fig_3}
\end{figure*}
\section{Discussion}\label{sec:disc}
\subsection{Comparison with numerical codes}
A comparison between the $R_{\rm CO}$ values for our reference cases obtained with the analytical solutions and with the numerical codes \texttt{DISKPOP} and \texttt{RADMC3D} from \citetalias{Toci21} is shown in Fig.~\ref{fig:fig_2}. In their work, the authors studied the time evolution of $R_{\rm CO}$ in a sample of viscously evolving discs, including CO freeze-out and CO photo-dissociation. To compare the numerical predictions with the results from observations, the authors evaluated the gas radii of the synthetic models as the radius which enclose the 68$\%$ and 90$\%$ of the CO flux, $R_{\rm CO, 68\%}$ and $R_{\rm CO, 90\%}$, while the analytical $R_{\rm CO}$, by definition, enclose 100$\%$ of the CO flux. We will firstly compare our analytic solution $R_{\rm CO}$ with $R_{\rm CO, 90\%}$ values from \citetalias{Toci21} and we will later discuss the implication of considering different values of the enclosed fluxes.
Note that, in order to compare our analytical prediction with the literature results, we selected the same initial conditions and we focused on the evolution between 0 and 3 Myr. We also fix the value of the stellar mass to 1 M$_\odot$ for simplicity, as in \citetalias{Toci21}. We are aware that the stellar mass plays a pivotal role in determining the initial conditions of discs (see e.g. \citealt{Somigliana22}), the gas and dust disc sizes, and disc evolution in general, but we defer to a future study on the effect of stellar mass.

Fig.~\ref{fig:fig_2} shows a comparison between the analytically evaluated $R_{\rm CO}$ of Fig.~\ref{fig:fig_1}.a (discs with $\gamma=1$, M$_0=0.1$ M$_\odot$, $R_c=10$ au and $\alpha=10^{-2},10^{-3},10^{-4}$, solid line) and values numerically obtained from models with the same intial conditions in \citetalias{Toci21} (dotted lines). Clearly, the trend of the numerical solutions are nicely recovered in the analytical solutions, especially for low and moderate values of the viscosity ($\alpha=10^{-3},10^{-4}$), but the analytical values of  $R_{\rm CO}$ are larger by a factor 2 with respect to the ones obtained with numerical codes (for example, the reference case has $R_{\rm CO}\sim$ 150 au and 70 au at $t=0.1$ Myr respectively, and the discrepancy remains amost fixed untill 3 Myr). The fact that the trends are recovered but the values are too large motivated us to try to better understand this discrepancy.

A first tentative way to reduce the analytical $R_{\rm CO}$ is by considering that they, by definition, enclose 100$\%$ of the $^{12}$CO flux rather 90$\%$.  Under the simplifying assumption that $^{12}$CO emission is optically thick at all radii the $^{12}$CO intensity profile follows the radial temperature profile, which itself is well described by a power-law $T(R)\sim R^{-\beta}$. It is then straightforward to show (see e.g., App. F in \citealt{Trapman19}) that $R_{\rm CO,90\%} = \lambda R_{\rm CO}$, where $\lambda =0.9^{1/(2-\beta)}$. However, for typical values $(\beta\approx 0.5)$ the effect of this is small ($\lambda \sim 0.93$),  $R_{\rm CO,90\%} \approx 0.93 R_{\rm CO}$. 
Fig.~\ref{fig:fig_2} also shows how the large discrepancy between the analytical solution and the numerical outputs cannot be solved if we simply consider also this numerical factor, which is a minor correction.

The only free parameter in the analytical solution is the relative $^{12}$CO abundance value $\xi_{\rm CO}$ in the definition of $\Sigma_{\rm crit}$. In our previous analysis we have assumed the standard $^{12}$CO abundance, $\xi_{\rm CO}=10^{-4}$. 
However, the value of $\xi_{\rm CO}$ is expected to be lower, as the region probed by $R_{\rm CO}$ is beyond the CO snowline. In the outermost, cold part of the disc $^{12}$CO can be removed from the gas phase and freeze-out onto dust grains, thus lowering $\xi_{\rm CO}$ (see \citealt{Pinte18} for a visualization of the CO snow surface in IM Lup).
In addition, recent observations have found several discs with a CO underabundancy by a factor 10-100 (e.g., \citealt{Favre2013,Cleeves2016} or \citealt{Miotello22} for a review), commonly interpreted as due to chemical processing of CO into other species.

\subsection{Effect of CO abundance}

A lower value of $\xi_{\rm CO}$ in the outer part of the discs could lead to a smaller observed gas disc sizes: the analytical solution of $R_{\rm CO}$ is sensitive to the $^{12}$CO content in the outer part of the disc. It is worth mentioning that this measure does not probe the overall abundance of $^{12}$CO in the warmer, inner part of the disc (where most of the mass is), which may be different. As the effect of CO freeze-out is also included in the analysis of \citetalias{Toci21}, a different $^{12}$CO abundance in the outer part of the disc can be responsible for the discrepancy between the two $R_{\rm CO}$ values. We then set the value of  $\xi_{\rm CO}=10^{-6}$, and we evaluate $R_{\rm CO}$ for the same set of initial conditions of Fig.~\ref{fig:fig_1}. We compare in Fig.~\ref{fig:fig_3} the values or $R_{\rm CO}$ obtained with the two different $^{12}$CO abundances ($\xi_{\rm CO}=10^{-4}$ and $10^{-6}$) and the results from \citetalias{Toci21}. As Fig.~\ref{fig:fig_3} shows, reducing the abundance of $^{12}$CO by 2 orders of magnitude can solve the discrepancy between the analytical and the numerical outcomes, with a few exceptions (very large discs or very low values of the disc mass), where an even higher $^{12}$CO depletion is needed. A lower value of $\xi_{\rm CO}$ deeply affects also the timescales on which the value of $R_{\rm CO}$ reaches its maximum, $t_{\rm inv}$. From eq.~\ref{eq:inv_time_reduced} we can see that for our reference case $t_{\rm inv}$ is reduced by a factor $100^{-2/5}$, $t_{\rm inv}\sim 21$ Myr. It is then possible to cross $t_{\rm inv}$ for initially low mass, large disc or very high values of the viscosity on timescales shorter than 10 Myr. For these discs, after reaching the maximum value, $R_{\rm CO}$ will \textit{decrease} with time even in the viscous scenario.

We can conclude that, once the effect of $^{12}$CO freeze out is properly calibrated in the models, the analytic solution can be an useful tool to evaluate  $R_{\rm CO}$ in population synthesis models of discs without using radiative transfer or thermochemical models, very expensive in computational times and resources.

Other, more complex processes can be at play, such as the chemical conversion of CO into other molecules on the dust grains or the dependence of the disc parameters and the CO abundance from the stellar mass. We defer to a future study a more quantitative analysis on these processes.

\section{Conclusions}\label{sec:conclusions}

In this Letter we derived an analytical solution for the time evolution of disc radii in protoplanetary discs using the flux emitted by $^{12}$CO as a tracer, $R_{\rm CO}$, assuming viscous evolution and that the $^{12}$CO radius is defined as the radius where the disc surface density is equal to the threshold for CO photo-dissociation. Our conclusions are:
\begin{itemize}
    \item This solution well describes the behaviour of previous results in the literature (obtained through numerical models including CO photo-dissociation and CO-freeze out) only if we adapt the value of the $^{12}$CO abundance to $\xi_{\rm CO}=10^{-6}$ at the location of $^{12}$CO photo-dissociation, two orders of magnitude lower than the standard assumption. This implies an efficient removal of CO molecules in the outer part of the disc, frozen onto grains.
    \item The solution implies a non-monotonic evolution of $R_{\rm CO}$, which initially expands due to viscous spreading, but later retreats. The inversion time, when we fix the $^{12}$CO abundance to $\xi_{\rm CO}=10^{-6}$ is $\sim$ 10 Myr, and might will be reached for populations of discs observable with ALMA. If the standard value of $\xi_{\rm CO}=10^{-4}$ is used, instead, the inversion time becomes much longer than the age of typical protostellar discs.
     \item Once appropriately calibrated, analytical solutions are a powerful tool to reduce the computational time needed to evaluate disc sizes in populations of discs. 
\end{itemize}

\section*{Acknowledgements}

The authors thank the reviewer Alexander Cridland for his thoughtful comments towards improving our manuscript. CT thanks Anna Miotello for her help and the helpful discussions. This work has received funding from the European Union’s Horizon 2020 research and innovation programme under the Marie Sklodowska Curie grant agreement N. 823823 (DUSTBUSTERS RISE project). GR acknowledges support from the Netherlands Organisation for Scientific Research (NWO, program number 016.Veni.192.233), from an STFC Ernest Rutherford Fellowship (grant number ST/T003855/1). This work is funded by the European Union under the European Union’s Horizon Europe Research $\&$ Innovation Programme grant No. 101039651 (DiscEvol). Views and opinions expressed are however those of the author(s) only and do not necessarily reflect those of the European Union or the European Research Council. Neither the European Union nor the granting authority can be held responsible for them.

\section*{Data Availability}

No new data were generated or analysed in support of this research.



\bibliographystyle{mnras}
\bibliography{mnras_template} 








\bsp	
\label{lastpage}
\end{document}